\documentclass[11pt]{article}
\usepackage{amsmath}
\usepackage{amsfonts}
\usepackage{amssymb}
\usepackage{graphicx}
\usepackage{subfigure}

\def\bea{\begin{eqnarray}}
\def\eea{\end{eqnarray}}

\begin{document}
\begin{center}
\LARGE {\bf Mixing-time and large-decoherence in continuous-time
quantum walks on one-dimension regular networks }
\end{center}
\begin{center}
{\bf  R. Radgohar$^a$ {\footnote {E-mail: r.radgohar@gmail.com}, S. Salimi$^a$ {\footnote {Corresponding author,
 E-mail: shsalimi@uok.ac.ir}}}}\\
 {$^a$\it Faculty of Science,  Department of Physics, University of Kurdistan, Pasdaran Ave., Sanandaj, Iran} \\
 \end{center}
\vskip 3cm
\begin{center}
{\bf{Abstract}}
\end{center}
In this paper, we study mixing and large decoherence in
continuous-time quantum walks on one dimensional regular networks,
which are constructed by connecting each node to its $2l$ nearest
neighbors($l$ on either side). In our investigation, the nodes of
network are represented by a set of identical tunnel-coupled quantum
dots in which decoherence is induced by continuous monitoring of
each quantum dot with nearby point contact detector. To formulate
the decoherent CTQWs, we use Gurvitz model and then calculate
probability distribution and the bounds of instantaneous and average
mixing times. We show that the mixing times are linearly
proportional to the decoherence rate. Moreover, adding links to
cycle network, in appearance of large decoherence, decreases the
mixing times.

\newpage

\section{Introduction}
Quantum walks are the quantum counterpart of random walks,
and were recently studied in the context of quantum information
because of their prominent role in the design of quantum
algorithms. Quantum walks were formulated in studies involving the
dynamics of quantum diffusion~\cite{FLS}, but the analysis of
quantum walks for use in quantum algorithms was first done by Farhi
and Gutmann~\cite{EFSG}. {Depending on the way the evolution
operator is defined, quantum walks can be either discrete-time
quantum walks(DTQWs)~\cite{ADZ} or continuous-time quantum
walks(CTQWs)~\cite{EFSG}. In the CTQW, one can directly define the
walk on the position space, whereas in the DTQW, it is necessary to
introduce a quantum coin operation to define the direction in which
the particle has to move. In recent years, many articles have
studied the dynamics of the quantum walks on networks. For example,
the DTQW has been studied in~\cite{NK,HKTB,MB,CMC} and the CTQW has
been considered in \cite{Gottlieb,tamon,SSS,xp1,
SJDP,SSQG,SSOG,saja,NKT,XPXR1,NKUS}.} One of the most important
quantities have been defined for quantum walks analogous to random
walks is mixing time. To introduce the mixing time, we refer to
computer science. In many computational problems, the best solution
can be found if we are able to sample from a well-chosen sampling
distribution. This can be provided by mapping the uniform
distribution into the desired one~\cite{DHS}. So, the behavior of
many algorithms that use quantum(random) walks depends on the time
it takes the walk to approach its uniform distribution, which is
called the mixing time. In other words, the quantum(classical)
algorithms are efficient if quantum(random) walks approach the
uniform distribution fast. Since any practical implementation scheme
of quantum walks must deal with decoherence, the natural question
rises is: what parameters affect the mixing time of decoherent
quantum walks. Several investigations on the decoherent quantum
walks were given in~\cite{Kendon, FWS, AR, RSAAD, KTNC, SRLS, SRLL}.
Also, the mixing time of decoherent CTQWs on cycle graphs has been
studied in~\cite{FST}. In that work, the authors proved that the
mixing time, for small rates of decoherence, improves linearly with
decoherence, whereas for large rates of decoherence, deteriorates
linearly towards the classical limit. But experimental
implementation of quantum walks is done by physical systems
including ground state atoms\cite{Dur} and Rydberg atoms\cite{cote}.
Some of the physical systems must deal with electromagnetism
interactions extending to long distances. For example, the clouds of
ultra cold Rydberg atoms assembled in a chain over which an exciton
migrates, the trapping of the exciton occurs \clearpage
\begin{figure}[h]
\vspace{2cm}\hspace{1.8cm}\includegraphics{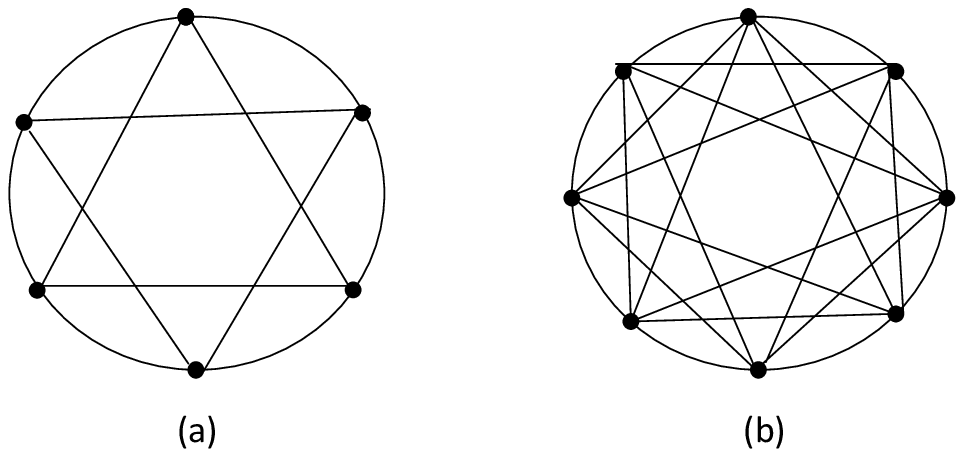}
 \vspace{-1cm}\caption{1D regular networks with $N=6$, $l=2$ (a) and  $N=8$ and $l=3$ (b). }
\end{figure}
at the ends of the chain. Therefore, we must take into account
long-range interactions by adding links to cycle and then study the
effects of these links on the CTQWs in the appearance of
decoherence. In this article, we focus on one dimensional ($1D$)
regular networks as generalized cycle networks having additional
links and show that for large rates of decoherence, the bounds of
instantaneous and average mixing times are proportional to
decoherence parameter but decrease with increasing additional links
(the problem for small decoherence is studied in~\cite{SRR}).
\\Our paper is structured as follows: We describe the structure of $1D$ regular
network in Sec. 2. Sec. 3 provides a brief summary of the main
concepts and of the formulae concerning CTQWs and give the exact
solutions to the transition probabilities on $1D$ regular network.
Sec. 4 presents the decoherent CTQWs on $1D$ regular network. We
assume that the decoherence rate is large and calculate the
probability distribution in Sec. 5.  In Sec. 6, we obtain the lower
and upper bounds of instantaneous and average mixing times.
Conclusions and discussions are given in the last part, Sec. 7.
\section{Structure of $1D$ regular network}
$1D$ networks are composed of a cycle graph of $N$ nodes in
which every node is connected to its $2l$ nearest neighbors($l$ on
either side)~\cite{XPXR}, where $N$ is the network size and $l$ is
the interaction parameter, i.e. all the two nodes whose distance is
smaller than or equal to $l$ are connected by additional bonds.
Figs. 1(a) , (b) show sketches of $1D$ regular networks with
$N=6,l=2$ and $N=8,l=3$, respectively. These networks provide a good
model to study various coupled dynamical systems, including
biological oscillators~\cite{HSIS}, Josephson junction
arrays~\cite{KW}, synchronization~\cite{IBBH}, small-world
networks~\cite{DWHS} and many other self-organizing systems.
\section{CTQWs on $1D$ regular network}
Every network can be considered as a graph made up of nodes and
algebraically described by the so-called adjacency matrix
$A=(A_{ij})$, which is a discrete version of the Laplace operator.
The non-diagonal elements $A_{ij}$ equal 1 if nodes $i$ and $j$ are
connected by a bond and 0 otherwise. The connectivity of node $i$
can be calculated as a sum of matrix elements
$z_{i}=\sum_{j}A_{ij}$. The Laplacian operator is then defined as
$L=Z-A$, where $Z$ is the diagonal matrix given by
$Z_{ik}=z_{i}\delta_{ik}$. It is worth underlining that, being
symmetric and non-negative definite, $L$ can generate both
probability conserving Markov process and unitary process. Thus, the
Laplacian operator can work both as a classical transfer operator
and as a tight-binding Hamiltonian of quantum transport
process~\cite{CG, MB5, V}.
\\The continuous-time random walks(CTRWs) are described by the
following Master equation~\cite{MW}:
\begin{eqnarray}\label{1}
\frac{d}{dt}p_{k,j}(t)=\sum_{l=1}^{N}T_{kl}p_{l,j}(t),
   \end{eqnarray}
being $p_{k,j}(t)$ the conditional probability that the walker is on
node $k$ at time $t$ when it started from node $j$. If the walk is
symmetric with a site-independent transmission rate $\gamma$, then
the transfer matrix $T$ is simply related to the Laplacian operator
through $T=-\gamma L$(in the following we set $\gamma=1$).
\\The quantum-mechanical extension of the CTRW is called
continuous-time quantum walk(CTQW). The CTQWs are obtained by
identifying the Hamiltonian of the system with the classical
transfer matrix, $H=-T$~\cite{EFSG, MB5, CFG}. The states
$|j\rangle$, representing the walker localized at the node $j$, span
the whole accessible Hilbert space and also provide an orthonormal
basis set. In these basis the Schr\"{o}dinger equation is
\begin{eqnarray}\label{2}
i\frac{d}{dt}|k\rangle=H|k\rangle,
   \end{eqnarray}
where we set $m=1$ and $\hbar=1$. The time evolution of state
$|j\rangle$ starting at time 0 is given by
$|j,t\rangle=U(t)|j\rangle$, where $U(t)=\exp[-iHt]$ is the
quantum-mechanical time evolution operator. Therefore, the behaviour
of the walker can be described by the transition amplitude
$\alpha_{k,j}(t)$ from state $|j\rangle$ to state $|k\rangle$, which
is
\begin{eqnarray}\label{3}
\alpha_{k,j}(t)=\langle k|e^{-iHt}|j\rangle.
   \end{eqnarray}

From Eq. (2), the $\alpha_{k,j}(t)$ obeys the following
Schr\"{o}dinger equation:
\begin{eqnarray}\label{4}
\frac{d}{dt}\alpha_{k,j}(t)=-i\sum_{l=1}^{N}H_{kl}\alpha_{l,j}(t).
   \end{eqnarray}
Note that the squared magnitude of transition amplitude provides the
quantum-mechanical transition probability
$\pi_{k,j}(t)=|\alpha_{k,j}(t)|^{2}$.
\\To get the exact solution of Eqs. (1) and (4), all the eigenvalues
and eigenvectors of the transfer operator and Hamiltonian are
required. We denote the $\emph{n}$th eigenvalue and eigenvector of
$H$ by $E_{n}$ and $|q_{n}\rangle$, respectively. Now, by using the
formal solution, the classical probability is given by
\begin{eqnarray}\label{5}
p_{k,j}(t)= {\langle k| e^{Tt}|j\rangle= \langle k|
e^{-Ht}|j\rangle}=\sum_{n=1}^{N}e^{-tE_{n}}\langle
k|q_{n}\rangle\langle q_{n}|j\rangle,
   \end{eqnarray}
and the quantum-mechanical transition probability can write as
\begin{eqnarray}\label{6}
\pi_{k,j}(t)=\sum_{n,l=1}^{N}e^{-it(E_{n}-E_{l})}\langle
k|q_{n}\rangle\langle q_{n}|j\rangle\langle q_{l}|k\rangle\langle
j|q_{l}\rangle,
   \end{eqnarray}
In the following, we focus on $1D$ regular networks and study CTQWs
on them. The Hamiltonian of the system is given by~\cite{SRR}
\begin{eqnarray}\label{7}
H_{ij}=\left\{
         \begin{array}{ll}
           -2l, & \hbox{if $i=j$;} \\
           1, & \hbox{if $i=j\pm m, m\in[1,l]$;} \\
           0, & \hbox{Otherwise.}
         \end{array}
       \right.
   \end{eqnarray}
This Hamiltonian acting on the state $|j\rangle$ can be written as
\begin{eqnarray}\label{8}
H|j\rangle=-(2l+1)|j\rangle+\sum_{m=-l}^{l}|j+m\rangle.
   \end{eqnarray}
which is the discrete version of the Hamiltonian for a free particle
moving on a lattice. It is well known in solid state physics that
the solutions of the Schr\"{o}dinger equation for a particle moving
freely in a regular potential are Bloch functions~\cite{OMAB, JMZ}.
We denote the Bloch states by $|\Phi_{n}\rangle$ and then the time
independent Schrodinger equation can be written as
\begin{eqnarray}\label{9}
H|\Phi_{n}\rangle=E_{n}|\Phi_{n}\rangle.
   \end{eqnarray}
The Bloch state $|\Phi_{n}\rangle$ can be expressed as a linear
combination of the states $|j\rangle$ localized at nodes $j$,
\begin{eqnarray}\label{10}
   |\Phi_{n}\rangle=\frac{1}{\sqrt{N}}\sum_{j=0}^{N-1}e^{-i\theta_{n}
   j}|j\rangle.
   \end{eqnarray}
Because of periodic boundary conditions, we have
$\Phi_{n}(N)=\Phi_{n}(0)$, where $\Phi_{n}(x)=\langle
x|\Phi_{n}\rangle$. This restricts the $\theta_{n}$-values to
$\theta_{n}=\frac{2\pi n}{N}$, where $n=0,1,\ldots ,N-1$. We set Eq.
(8) and (10) into (9) and obtain the eigenvalues of system
as~\cite{SRR}
\begin{eqnarray}\label{8}
   E_{n}=-2l+2\sum_{j=1}^{l}\cos(j\theta_{n}).
   \end{eqnarray}

Thus Eq. (5) and (6) can be written as
\begin{eqnarray}\label{8}
   p_{k,j}(t)=\frac{1}{N}\sum_{n}e^{-tE_{n}}e^{-i(k-j)\frac{2n\pi}{N}},
   \end{eqnarray}
\begin{eqnarray}\label{8}
   \pi_{k,j}(t)=\frac{1}{N^{2}}\sum_{n,l}e^{-it(E_{n}-E_{l})}e^{-i(k-j)(n-l)\frac{2\pi}{N}}.
   \end{eqnarray}

\section{The Decoherent CTQWs on $1D$ regular network}
In this section, we take into account the effects of decoherence in
quantum walks on $1D$ regular network. For this aim, we consider the
model of network in which decoherence is induced by continuous
monitoring of each network node with nearby point contact(PC)
detector. In this model, our assumptions are as follows: each node
is represented by a quantum dot which is continuously monitored by
an individual point contact(PC), the walks are performed by an
electron initially placed in one of the quantum dots, identical PCs
are placed far enough from QDs so that the tunneling between them is
negligible, Coulomb interaction between electrons in QD and PC is
taken into account and all electrons are spin-less
fermions~\cite{DSLF}. With help of gate-engineering techniques in
semiconductor heterostructures, the quantum dots cycle (the 1D
regular graphs are a kind of cycle graphs) with attached point
contacts can be constructed\cite{torre}. Such techniques allow
electronically to make  of quantum dots and point contact by placing
metal gates on the structure with a two-dimensional electron gas. By
changing the potential on gates one can assign area of
two-dimensional electron gas \clearpage
\begin{figure}[h] \vspace{5cm} \includegraphics{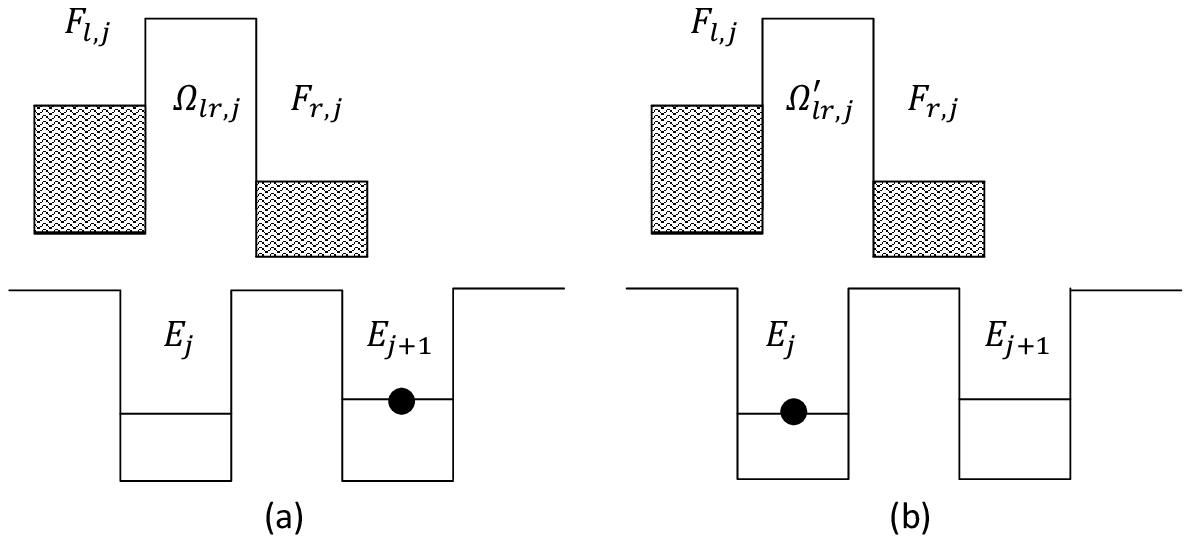} \vspace{-1cm}
\caption{Fig. 2(a) shows point contact detector $j$ monitoring the
electron in dot $j+1$ and Fig. 2(b) shows point contact detector $j$
when electron is placed in dot $j$. Source and drain reservoirs are
kept at zero temperature with chemical potentials $F_{l,j}$ and
$F_{r,j}$, respectively. $E_{j}$ is the on-site node energy.
$\Omega_{lr,j}$ and $\Omega^{\prime}_{lr,j}$ are transmission
probability of the detector for empty dot and for the occupied dot.}
\end{figure}which creates the necessary confinement
profile~\cite{DSLF}. In Ref.~\cite{Pioro}, the simplest example of
such structure which contains two quantum dots was experimentally
investigated. Such a set up is shown schematically in Fig. 2, where
detector $j$ is represented by a barrier, connected with two
reservoirs at the potentials $F_{l,j}$ and $F_{r,j}$. The
transmission probability of the barrier varies from $\Omega_{lr,j}$
to $\Omega^{\prime}_{lr,j}$, depending on whether or not the quantum
dot is occupied by an electron. In the following, we want to write
the Hamiltonian for the entire system. We consider simple
continuous-time quantum walks are defined over an undirected graph
with $N$ nodes in which each node corresponds by an integer
$j\in[0,N-1]$ and a quantum state $|j\rangle$. These walks can be
well described by the following Hamiltonian~\cite{DHS, APS}:
\begin{eqnarray}\label{13}
\begin{array}{cc}
  H_{s}= & \displaystyle\sum_{ij}\Delta_{ij}(t)(\hat{c}^{\dag}_{i}\hat{c}_{j}+\hat{c}_{i}\hat{c}^{\dag}_{j})-
\displaystyle\sum_{j}E_{j}(t)\hat{c}^{\dag}_{j}\hat{c}_{j}, \\
  & \\
   =\hspace{-.5cm}&\displaystyle\sum_{ij}\Delta_{ij}(t)(|i\rangle\langle j|+|j\rangle\langle
   i|)+\displaystyle\sum_{j}E_{j}(t)|j\rangle\langle j|.
\end{array}
     \end{eqnarray}
where $\hat{c}^{\dag}_{i}(\hat{c}_{j})$ are creation (annihilation)
operators such that $\hat{c}^{\dag}_{i}\hat{c}_{j}$ acting on a
state at node $j$ brings it to node $i$. So, state $|j\rangle$
denoting the state of 'particle' at node $j$ can be obtained by
acting $c^{\dag}_{j}$ on the ground state $|0\rangle$. The two terms
correspond to a hopping term with amplitudes $\triangle_{ij}(t)$
between nodes, and on-site node energies $E_{j}(t)$, both of which
can depend on time. For the sake of simplicity, we drop all the
on-site energies(i.e. $E_j=0, \forall j$) and assume that hopping
amplitudes($\triangle_{ij}(t)$) between connected sites to be
constant. Also, we renormalize the time, so that it becomes
dimensionless~\cite{DSLF}. Thus, Eq. (14) for $1D$ regular networks
is written as
\begin{eqnarray}\label{14}
  H_{s}=\frac{1}{4}\sum_{j=0}^{N-1}\sum_{z=1}^{l}(\hat{c}^{\dagger}_{j+z}\hat{c}_{j}+\hat{c}^{\dag}_{j}c_{j+z}).
\end{eqnarray}
Now, we study the electron transport process in point contact $j$.
The point contact is considered as a barrier, separated two
reservoirs(the source and drain). All the levels in the source and
drain are initially filled up to the Fermi energies, which is called
the vacuum state $|0\rangle$. Thus, the Hamiltonian of $j$-th point
contact can be written as
\begin{eqnarray}\label{15}
  H_{pc,j}=\sum_{l}E_{l,j}\hat{a}^{\dagger}_{l,j}\hat{a}_{l,j}+\sum_{r}E_{r,j}\hat{a}^{\dagger}_{r,j}\hat{a}_{r,j}+
  \sum_{l,r}\Omega_{lr,j}(\hat{a}^{\dagger}_{l,j}\hat{a}_{r,j}+\hat{a}^{\dagger}_{r,j}\hat{a}_{l,j}),
\end{eqnarray}
where $\hat{a}^{\dagger}_{l,j}(\hat{a}_{l,j})$ and
$\hat{a}^{\dagger}_{r,j}(\hat{a}_{r,j})$ are the
creation(annihilation) operators in the left and right reservoirs
respectively, and $\Omega_{lr,j}$ is the hopping amplitude between
the states $E_{l,j}$ and $E_{r,j}$ in the right and left reservoirs.
\\The interaction between the detector and the measured system is
described by $H_{int}$. The presence of an electron in the left dot
results in an effective increase of the point contact
barrier($\acute{\Omega_{lr,j}}=\Omega_{lr,j}+\delta\Omega_{lr,j}$).
Therefore, the interaction Hamiltonian can be written as
\begin{eqnarray}\label{16}
  H_{int}=\sum_{l,r}\delta\Omega_{lr,j}\hat{c}_{j}^{\dagger}\hat{c}_{j}(\hat{a}_{l,j}^{\dagger}\hat{a}_{r,j}+\hat{a}^{\dagger}_{r,j}\hat{a}_{l,j}).
\end{eqnarray}
For simplicity, we assume that the hoping amplitudes are weakly
dependent on states $E_{l,j}$ and $E_{r,j}$, so that
$\Omega_{lr,j}=\bar{\Omega}$,
$\delta\Omega_{lr,j}=\delta\bar{\Omega}$ and
$F_{l,j}(F_{r,j})=\bar{F_{l}}(\bar{F_{r}})$. The decoherence rate
$\Gamma$ can be written as
$\Gamma=\delta\bar{\Omega}^2(\bar{F}_{r}-\bar{F}_{l})^{2}f_{S}f_{D}$
where $f_{S}$ and $f_{D}$ are density of states in source and
reservoirs, respectively~\cite{DSLF}. Gurvitz in~\cite{G} showed
that the appearance of decoherence leads to the collapse of the
density matrix into the statistical mixture in the course of the
measurement processes, thus the evolution of the reduced density
matrix traced over all states of source and drain electrons is given
by Bloch-type rate equations. The time dependent non-unitary
evolution of reduced density matrix
$\rho(t)=|\Phi(t)\rangle\langle\Phi(t)|$ in the Gurvitz model is
given by~\cite{SRR}

\begin{eqnarray}\label{17}
\begin{array}{cc}
  \frac{d}{dt}\rho_{j,k}(t)= & -\frac{i}{4}[\displaystyle\sum_{m=-l}^{l}(\rho_{j+m,k}-\rho_{j,k+m})]-\Gamma(1-\delta_{j,k})\rho_{j,k}\hspace{2.4cm} \\
   &  \\
  =\hspace{-1.5cm} &
  -\frac{i}{4}[\displaystyle\sum_{m=1}^{l}(\rho_{j+m,k}-\rho_{j,k+m}+\rho_{j-m,k}-\rho_{j,k-m})]-\Gamma(1-\delta_{j,k})\rho_{j,k}.
\end{array}
\end{eqnarray}

\section{Large Decoherence}
In~\cite{SRR}, authors studied the effect of small
decoherence($\Gamma N\ll 1$) in CTQWs on $1D$ regular networks. They
showed that the instantaneous mixing time upper bound and the
average time mixing lower bound are independent of parameter
$l(l\geq 2)$, but are proportional to the inverse of decoherence
rate. Here, we want to study CTQWs on these networks in appearance
of the large rates of decoherence. For this purpose, we use Gurvitz
model and focus on the elements of matrix $\rho(t)$. Based on the
initial conditions, the non-zero elements appear only along the
major diagonal. Firstly, we rewrite Eq. (18) for the elements of
major diagonal and also minor diagonals whose distances of the major
diagonal are lesser than $l$. By dropping terms that are smaller
than $1/\Gamma$, we have

\begin{eqnarray}\label{17}
\left\{
  \begin{array}{ll}
    \rho^{\prime}_{j,j}(t)=-\frac{i}{4}[\displaystyle\sum_{m=1}^{l}(\rho_{j+m,j}-\rho_{j,j+m}+\rho_{j-m,j}-\rho_{j,j-m})],  \\
     & \\
    \rho^{\prime}_{j,j+1}(t)=-\frac{i}{4}[\rho_{j+1,j+1}-\rho_{j,j}]-\Gamma \rho_{j,j+1},  \\
     & \\
    \rho^{\prime}_{j,j+2}(t)=-\frac{i}{4}[\rho_{j+2,j+2}-\rho_{j,j}]-\Gamma \rho_{j,j+2},\\
              \vdots\\
        \rho^{\prime}_{j,j+l}(t)=-\frac{i}{4}[\rho_{j+l,j+l}-\rho_{j,j}]-\Gamma \rho_{j,j+l},
  \end{array}
\right.
\end{eqnarray}
For the simplicity sake, we use the following definitions
\begin{eqnarray}\label{17}
  \begin{array}{cc}
    a_{j}=\rho_{j,j}\\
    d_{j}=\rho_{j,j+1}-\rho_{j+1,j},  \\
    f_{j}=\rho_{j,j+2}-\rho_{j+2,j}, \\
        \vdots &  \\
    q_{j}=\rho_{j,j+l}-\rho_{j+l,j}. &
  \end{array}
\end{eqnarray}
Then the above difference equation system can be written as
\begin{eqnarray}\label{17}
\left\{
  \begin{array}{ll}
   a^{\prime}_{j}=-\frac{i}{4}[-d_{j}+d_{j-1}-f_{j}+f_{j-2}+\cdots-q_{j}+q_{j-l}],\\
      \\
     d^{\prime}_{j}=-\frac{i}{2}[a_{j+1}-a_{j}]-\Gamma d_{j},\\
     &  \\
    f^{\prime}_{j}=-\frac{i}{2}[a_{j+2}-a_{j}]-\Gamma f_{j},\\
          \vdots  \\
    q^{\prime}_{j}=-\frac{i}{2}[a_{j+l}-a_{j}]-\Gamma q_{j},\\
  \end{array}
\right.
\end{eqnarray}
Differentiation of the above equation gives
\begin{eqnarray}\label{17}
\left\{
  \begin{array}{ll}
   a^{\prime \prime}_{j}=-\frac{i}{4}[-d^{\prime}_{j}+d^{\prime}_{j-1}-f^{\prime}_{j}+f^{\prime}_{j-2}+\cdots-q^{\prime}_{j}+q^{\prime}_{j-l}],\\
      \\
     d^{\prime \prime}_{j}=-\frac{i}{2}[a^{\prime}_{j+1}-a^{\prime}_{j}]-\Gamma d^{\prime}_{j},\\
     &  \\
    f^{\prime \prime}_{j}=-\frac{i}{2}[a^{\prime}_{j+2}-a^{\prime}_{j}]-\Gamma f^{\prime}_{j},\\
          \vdots  \\
    q^{\prime \prime}_{j}=-\frac{i}{2}[a^{\prime}_{j+l}-a^{\prime}_{j}]-\Gamma q^{\prime}_{j},\\
  \end{array}
\right.
\end{eqnarray}
We can guess the following solutions for the above equations

\begin{eqnarray}\label{17}
\begin{array}{ccc}
  a_{j}=\displaystyle\sum_{k=0}^{N-1}A_{k}e^{\frac{2\pi ijk}{N}}e^{-\gamma_{k}t}, &  & d_{j}=\displaystyle\sum_{k=0}^{N-1}D_{k}e^{\frac{2\pi ijk}{N}}e^{-\gamma_{k}t}, \\
   &  &  \\
  f_{j}=\displaystyle\sum_{k=0}^{N-1}F_{k}e^{\frac{2\pi ijk}{N}}e^{-\gamma_{k}t}, & \cdots, & q_{j}=\displaystyle\sum_{k=0}^{N-1}Q_{k}e^{\frac{2\pi ijk}{N}}e^{-\gamma_{k}t},
\end{array}
\end{eqnarray}

in which $\gamma_{k}$, $A_{k}$, $D_{k}$,$\cdots$ and $Q_{k}$ are the
unknown quantities. Then, we set these solutions into Eqs. (22) and
get
\begin{eqnarray}\label{17}
\left\{
  \begin{array}{ll}
    \gamma_{k}A_{k}+\frac{i}{4}[D_{k}(1-e^{\frac{-2\pi
ik}{N}})+F_{k}(1-e^{\frac{-4\pi
ik}{N}})+\cdots+Q_{k}(1-e^{\frac{-2\pi ikl}{N}})]=0\\
&\\
    A_{k}[\frac{i}{2}(-e^{\frac{2\pi ik}{N}}+1)]+D_{k}(\gamma_{k}-\Gamma)=0,\\
&\\
    A_{k}[\frac{i}{2}(-e^{\frac{4\pi ik}{N}}+1)]+F_{k}(\gamma_{k}-\Gamma)=0, \\
        \vdots  \\
    A_{k}[\frac{i}{2}(-e^{\frac{2l\pi ik}{N}}+1)]+Q_{k}(\gamma_{k}-\Gamma)=0,
  \end{array}
\right.
\end{eqnarray}
It is evident that there are nontrivial solutions for the set of
equations if the determinant of the coefficients matrix is zero.
\begin{eqnarray}\label{10}
  (\gamma_{k}-\Gamma)^{(l-1)}[\gamma_{k}(\gamma_{k}-\Gamma)+\frac{1}{2}(\sin^{2}(\frac{\pi k}{N})+\sin^{2}(\frac{2\pi
k}{N})+\cdots+\sin^{2}(\frac{l\pi k}{N}))]=0
   \end{eqnarray}
Therefore, four values for $\gamma_{k}$ are obtained as
\begin{eqnarray}\label{10}
\gamma_{k}=\left\{
  \begin{array}{ll}
    \gamma_{k,0}=0, \\
    \gamma_{k,1}=\Gamma, \\
    \gamma_{k,2}=\Gamma-\displaystyle\frac{1}{2\Gamma}\displaystyle\sum_{m=1}^{l}\sin^{2}(\frac{\pi km}{N}),  \\
         \gamma_{k,3}=\displaystyle\frac{1}{2\Gamma}\displaystyle\sum_{m=1}^{l}\sin^{2}(\frac{\pi km}{N}),
  \end{array}
\right.
   \end{eqnarray}
The general solutions of Eqs. (22) are

\begin{eqnarray}\label{27}
 \left\{
   \begin{array}{ll}
     a_{j}=\frac{1}{N}\displaystyle\sum_{k=0}^{N-1}\{A_{k,0}e^{-\gamma_{k,0}t}+A_{k,1}e^{-\gamma_{k,1}t}+A_{k,2}e^{-\gamma_{k,2}t}+A_{k,3}e^{-\gamma_{k,3}t}\}\omega^{jk},\\
      \\
     d_{j}=\frac{1}{N}\displaystyle\sum_{k=0}^{N-1}\{D_{k,0}e^{-\gamma_{k,0}t}+D_{k,1}e^{-\gamma_{k,1}t}+D_{k,2}e^{-\gamma_{k,2}t}+D_{k,3}e^{-\gamma_{k,3}t}\}\omega^{jk},\\
       \\
     f_{j}=\frac{1}{N}\displaystyle\sum_{k=0}^{N-1}\{F_{k,0}e^{-\gamma_{k,0}t}+F_{k,1}e^{-\gamma_{k,1}t}+F_{k,2}e^{-\gamma_{k,2}t}+F_{k,3}e^{-\gamma_{k,3}t}\}\omega^{jk},\\
           \vdots \\
     q_{j}=\frac{1}{N}\displaystyle\sum_{k=0}^{N-1}\{Q_{k,0}e^{-\gamma_{k,0}t}+F_{k,1}e^{-\gamma_{k,1}t}+F_{k,2}e^{-\gamma_{k,2}t}+F_{k,3}e^{-\gamma_{k,3}t}\}\omega^{jk},
   \end{array}
 \right.
   \end{eqnarray}

where $\omega=e^{\frac{2\pi i}{N}}$. By the initial condition
$a_{j}(0)=\delta_{j,0}$ and $d_{j},f_{j},...,q_{j}(0)=0$, for
$j=0,\cdots, N-1$, and replacing the constant coefficients into the
others, we get

\begin{eqnarray}\label{30}
\left\{
\begin{array}{ccc}
  A_{k,1}\simeq0, & A_{k,2}\simeq\frac{-1}{2\Gamma^{2}}\displaystyle\sum_{m=1}^{l}\sin^{2}(\frac{\pi km}{N}), & A_{k,3}=1,\\
\\
D_{k,1}\simeq0,&D_{k,2}\simeq\frac{i}{\Gamma}\sin(\frac{\pi
k}{N})\exp(\frac{i\pi k}{N}),&
D_{k,3}\simeq\frac{-i}{\Gamma}\sin(\frac{\pi k}{N})\exp(\frac{i\pi
k}{N}),\\
\\
F_{k,1}\simeq0,&F_{k,2}\simeq\frac{i}{\Gamma}\sin(\frac{2\pi
k}{N})\exp(\frac{2i\pi k}{N}),&
F_{k,3}\simeq\frac{-i}{\Gamma}\sin(\frac{2\pi k}{N})\exp(\frac{2i\pi
k}{N}),\\
\vdots\\
Q_{k,1}\simeq0,&Q_{k,2}\simeq\frac{i}{\Gamma}\sin(\frac{l\pi
k}{N})\exp(\frac{li\pi k}{N}),&
Q_{k,3}\simeq\frac{-i}{\Gamma}\sin(\frac{l\pi k}{N})\exp(\frac{li\pi
k}{N}).
\end{array}
\right.
   \end{eqnarray}
Note that the probability distribution $P(t)$ of the quantum walk is
specified by the diagonal elements of $\rho(t)$, that is
$P_{j}(t)=a_{j}(t)$. For our problem, this distribution reduces to
\begin{eqnarray}\label{3}
a_{j}(t)=\frac{1}{N}\displaystyle\sum_{k=0}^{N-1}\exp
[-\frac{t\sum_{m=1}^{l}\sin^{2}(\frac{\pi
km}{N})}{2\Gamma}]\omega^{jk}
   \end{eqnarray}
Since CTQWs are symmetric under time-inversion, the above
distribution does not converge to any constant value.
\section{Mixing time}
In this section, we discuss the rate of convergence to the above
probability distribution which can be expressed in terms of mixing
time. There are two distinct notations of mixing time in the
literature: the instantaneous mixing time and the average mixing
time. In the following, we give the definition of the two notations
of mixing time in the continuous-time quantum walks on graphs and
calculate them for our network.
\\
\\(a) \textbf{\textit{Instantaneous mixing time}}
\\The instantaneous mixing time focuses on particular times at which
the probability distribution is sufficiently close to the uniform
distribution~\cite{DHS}, i.e. $t_{ins}$ is the instantaneous mixing
time if
\begin{eqnarray}\label{35}
t_{ins}=min\{t :
\sum_{j=0}^{N-1}\|P_{j}(t)-\frac{1}{N}\|\leq\epsilon\},
   \end{eqnarray}
where $||p-q||=\sum_{i}|p_{i}-q_{i}|$ denotes the total variation
distance between the distributions $p$ and $q$.
\\Adding a summation over $j$ to Eq. (29) gives us the total variation
distance required to get the mixing time:
\begin{eqnarray}\label{35}
\begin{array}{cc}
  \displaystyle\sum_{j=0}^{N-1}|a_{j}(t)-\frac{1}{N}| & =\displaystyle\sum_{j=0}^{N-1}|\frac{1}{N}\sum_{k=0}^{N-1}\exp[\frac{-t}{2\Gamma}\displaystyle\sum_{m=1}^{l}\sin^{2}(\frac{\pi
km}{N})]\omega^{jk}-\frac{1}{N}|\hspace{1.8cm} \\
   &  \\
   & =\displaystyle\sum_{j=0}^{N-1}|\frac{1}{N}\displaystyle\sum_{k=1}^{N-1}\exp[\frac{-t}{2\Gamma}\displaystyle\sum_{m=1}^{l}\sin^{2}(\frac{\pi
km}{N})]\cos(\frac{2\pi jk}{N})|\hspace{1.5cm}
\end{array}
   \end{eqnarray}
\textmd{\textbf{Lower bound}}
\\ To find the lower bound, we use
the term $j=0$ only
\begin{eqnarray}\label{35}
\begin{array}{cc}
  \displaystyle\sum_{j=0}^{N-1}|a_{j}(t)-\frac{1}{N}| & >|a_{0}(t)-\displaystyle\frac{1}{N}|=\displaystyle\frac{1}{N}|\displaystyle\sum_{k=1}^{N-1}\exp[\frac{-t}{2\Gamma}\displaystyle\sum_{m=1}^{l}\sin^{2}(\frac{\pi
km}{N})]|\hspace{1.8cm} \\
\end{array}
   \end{eqnarray}
then consider the terms $k=1,N-1$
   \begin{eqnarray}\label{36}
   \begin{array}{cc}
   \displaystyle\sum_{j=0}^{N-1}|a_{j}(t)-\frac{1}{N}| & >\displaystyle\frac{2}{N}e^{-\displaystyle\frac{t}{2\Gamma}\displaystyle\sum_{m=1}^{l}\sin^{2}(\frac{\pi
m}{N})}.
\end{array}
   \end{eqnarray}

It reaches $\epsilon$ at time $t_{ins,lower}$ when
\begin{eqnarray}\label{8}
\frac{2}{N}e^{-\displaystyle\frac{t_{ins,lower}}{2\Gamma}\displaystyle\sum_{m=1}^{l}\sin^{2}(\displaystyle\frac{\pi
m}{N})}=\epsilon.
   \end{eqnarray}
Finally, the instantaneous mixing time lower bound is obtained as
\begin{eqnarray}\label{8}
t_{ins,lower}=\frac{2\Gamma
}{\displaystyle\sum_{m=1}^{l}\sin^{2}(\frac{\pi
m}{N})}\ln(\frac{2}{N\epsilon}).
   \end{eqnarray}

Note that for $l=1$ and large $N\gg1$, we have
\begin{eqnarray}\label{8}
t_{ins,lower}=\frac{2\Gamma}{\sin^{2}(\frac{\pi}{N})}\ln(\frac{2}{N\epsilon})\simeq\frac{2\Gamma
N^{2}}{\pi^{2}}\ln(\frac{2}{N\epsilon})
   \end{eqnarray}
which is in agreement with~\cite{FST}'s result for cycle.
\\
\\\textmd{\textbf{Upper bound}}
\\An upper bound on the instantaneous mixing time can also be
derived. To do this, first we find an upper bound for Eq. (31) by
using the relation $|\cos(\frac{2\pi jk}{N})|<1$ for $k,j\neq0,N$.
 \begin{eqnarray}\label{35}
 \begin{array}{cc}
   \displaystyle\sum_{j=0}^{N-1}|a_{j}(t)-\frac{1}{N}|
   <\displaystyle\frac{1}{N}\displaystyle\sum_{j=0}^{N-1}\displaystyle\sum_{k=1}^{N-1}\exp[\frac{-t}{2\Gamma}\displaystyle\sum_{m=1}^{l}\sin^{2}(\frac{\pi
   km}{N})]\hspace{3.6cm}\\
  &\\
   < \displaystyle\frac{2}{N}\displaystyle\sum_{j=0}^{N-1}\displaystyle\sum_{k=1}^{[N/2]}\exp[\frac{-t}{2\Gamma}\displaystyle\sum_{m=1}^{l}\sin^{2}(\frac{\pi
   km}{N})]\hspace{0.9cm} \\
   \end{array}
   \end{eqnarray}
Since $\sin(x)>\frac{2x}{\pi}$ for $0<x<\frac{\pi}{2}$~\cite{MASH} ,
we have
\begin{eqnarray}\label{43}
\begin{array}{cc}
  \displaystyle\sum_{j=0}^{N-1}|a_{j}(t)-\frac{1}{N}|<
   \displaystyle\frac{2}{N}\displaystyle\sum_{j=0}^{N-1}\displaystyle\sum_{k=1}^{[N/2]}\exp[\frac{-t}{\Gamma}\displaystyle\sum_{m=1}^{l}\frac{2k^{2}}{N^{2}m^{2}}]\hspace{2.6cm} \\
     \end{array}
      \end{eqnarray}
   and by using $k^{2}\geq k$ when $k\geq 1$, we get
   \begin{eqnarray}\label{43}
\begin{array}{cc}
 \displaystyle\sum_{j=0}^{N-1}|a_{j}(t)-\frac{1}{N}|<
 &\displaystyle\frac{2}{N}\displaystyle\sum_{j=0}^{N-1}\displaystyle\sum_{k=1}^{\infty}\exp[\frac{-t}{\Gamma}\displaystyle\sum_{m=1}^{l}\frac{2k}{N^{2}m^{2}}]\hspace{3cm}
\end{array}
   \end{eqnarray}
 After some algebra, we obtain
\begin{eqnarray}\label{44}
\sum_{j=0}^{N-1}|a_{j}(t)-\frac{1}{N}|<\frac{2}{\exp[{\displaystyle\frac{t}{\Gamma}\displaystyle\sum_{m=1}^{l}\frac{2}{N^{2}m^{2}}}]-1}.
   \end{eqnarray}
According to the instantaneous mixing time definition(Eq.(30)):
\begin{eqnarray}\label{44}
\frac{2}{\exp[{\displaystyle\frac{t_{ins,upper}}{\Gamma}\displaystyle\sum_{m=1}^{l}\frac{2}{N^{2}m^{2}}}]-1}=\epsilon.
   \end{eqnarray}
Therefore, the upper bound of instantaneous mixing time is
\begin{eqnarray}\label{44}
t_{ins,upper}=\frac{\Gamma
N^{2}}{2\displaystyle\sum_{m=1}^{l}m^{-2}}\ln(\frac{2+\epsilon}{\epsilon}).
   \end{eqnarray}
Moreover, for $l=1$(cycle), we obtain
\begin{eqnarray}\label{44}
t_{ins,upper}=\frac{\Gamma N^{2}}{2}\ln(\frac{2+\epsilon}{\epsilon})
   \end{eqnarray}
which is the same result mentioned in~\cite{FST}.
\\
\\
(b) \textbf{\textit{Average mixing time}}
\\The average mixing time which is based on the time-averaged
probability distribution
\begin{eqnarray}\label{44}
\bar{P}(j,T)=\frac{1}{T}\int_{0}^{T}P(j,t)dt,
   \end{eqnarray}
is the time it takes the average distribution to be $\epsilon$-close
to the uniformly distributed~\cite{AJVK}, i.e. $t_{ave}$ is the
average mixing time if
\begin{eqnarray}\label{35}
t_{ave}=min\{t : \|\bar{P}(j,T)-\frac{1}{N}\|\leq\epsilon\}.
   \end{eqnarray}
where $||.||$ is the total variation distance mentioned in the
instantaneous mixing time. To get the bounds of average mixing time,
first we calculate $\bar{P}(j,T)$ by Eq. (29)

\begin{eqnarray}\label{44}
\begin{array}{cc}
  \bar{P}(j,T) &=\displaystyle\frac{1}{T}\displaystyle\int_{0}^{T}\displaystyle\frac{1}{N}\displaystyle\sum_{k=0}^{N-1}\exp[-\displaystyle\frac{t}{2\Gamma}\displaystyle\sum_{m=1}^{l}\sin^{2}(\frac{\pi
km}{N})]\omega^{jk}dt \\
   &  \\
   & =\displaystyle\frac{2\Gamma}{TN}\displaystyle\sum_{k=0}^{N-1}\displaystyle\frac{1-\exp[-\displaystyle\frac{T}{2\Gamma}\displaystyle\sum_{m=1}^{l}\sin^{2}(\frac{\pi km}{N})]\omega^{jk}}{\displaystyle\sum_{m=1}^{l}\sin^{2}(\frac{\pi km}{N})}
\end{array}
   \end{eqnarray}
The total variation distance between the uniform distribution and
the time-average distribution of the decoherent quantum walk is
given by

\begin{eqnarray}\label{44}
\begin{array}{cc}
  \displaystyle\sum_{j=0}^{N-1}|\bar{P}(j,T)-\frac{1}{N}|  & = \displaystyle\sum_{j=0}^{N-1}|\displaystyle\frac{2\Gamma}{TN}\displaystyle\sum_{k=0}^{N-1}\displaystyle\frac{1-\exp[-\displaystyle\frac{T}{2\Gamma}\displaystyle\sum_{m=1}^{l}\sin^{2}(\displaystyle\frac{\pi km}{N})]\omega^{jk}}{\displaystyle\sum_{m=1}^{l}\sin^{2}(\displaystyle\frac{\pi
km}{N})}-\displaystyle\frac{1}{N}| \\
   &  \\
   & =\displaystyle\sum_{j=0}^{N-1}|\displaystyle\frac{2\Gamma}{TN}\displaystyle\sum_{k=1}^{N-1}\displaystyle\frac{1-\exp[-\displaystyle\frac{T}{2\Gamma}\displaystyle\sum_{m=1}^{l}\sin^{2}(\displaystyle\frac{\pi km}{N})]\omega^{jk}}{\displaystyle\sum_{m=1}^{l}\sin^{2}(\displaystyle\frac{\pi km}{N})}| \\
\end{array}
\end{eqnarray}
\\
\\\textmd{\textbf{Lower bound}}
\\Now, we can find a lower bound for the average mixing time.
As with the lower bound of instantaneous mixing time, we use the
terms $j=0, k=1,N-1$.
\begin{eqnarray}\label{44}
\begin{array}{cc}
      \displaystyle\sum_{j=0}^{N-1}|\bar{P}(j,T)-\frac{1}{N}| &
        >\displaystyle\frac{4\Gamma}{TN}\displaystyle\frac{1-\exp[-\displaystyle\frac{T}{2\Gamma}\displaystyle\sum_{m=1}^{l}\sin^{2}(\displaystyle\frac{\pi
m}{N})]}{\displaystyle\sum_{m=1}^{l}\sin^{2}(\displaystyle\frac{\pi
m}{N})}
\end{array}
   \end{eqnarray}
\\Assume $T/\Gamma\gg 1$
(which is consistent with our other assumptions, $N\gg 1$ and
$\Gamma N\gg 1$, it requires $T\gg N$) which results in
\begin{eqnarray}\label{44}
\displaystyle\sum_{j=0}^{N-1}|\bar{P}(j,T)-\frac{1}{N}|>\displaystyle\frac{4\Gamma}{TN\displaystyle\sum_{m=1}^{l}\sin^{2}(\displaystyle\frac{\pi
m}{N})}
   \end{eqnarray}
Therefore, the lower bound of average mixing time is
\begin{eqnarray}\label{44}
t_{ave,lower}=\frac{4\Gamma}{\epsilon N
\sum_{m=1}^{l}\sin^{2}(\frac{\pi m}{N})}.
   \end{eqnarray}
   \\
\\\textmd{\textbf{Upper bound}}
\\An upper bound for the average mixing time can be obtained
by using the technique mentioned for the upper bound of
instantaneous mixing time. So, we have
\begin{eqnarray}\label{44}
\begin{array}{cc}
  \displaystyle\sum_{j=0}^{N-1}|\bar{P}(j,t)-\frac{1}{N}| &
    =\displaystyle\sum_{j=0}^{N-1}|\displaystyle\frac{2\Gamma}{TN}\displaystyle\sum_{k=1}^{N-1}\displaystyle\frac{1-\exp[-\displaystyle\frac{T}{2\Gamma}\displaystyle\sum_{m=1}^{l}\sin^{2}(\displaystyle\frac{\pi
km}{N})]\cos(\displaystyle\frac{2\pi
jk}{N})}{\displaystyle\sum_{m=1}^{l}\sin^{2}(\displaystyle\frac{\pi
km}{N})}| \\
   & \\
   & <\displaystyle\sum_{j=0}^{N-1}|\displaystyle\frac{2\Gamma}{TN}\displaystyle\sum_{k=1}^{N-1}\displaystyle\frac{1-\exp[-\displaystyle\frac{T}{2\Gamma}\displaystyle\sum_{m=1}^{l}\sin^{2}(\displaystyle\frac{\pi
km}{N})]}{\displaystyle\sum_{m=1}^{l}\sin^{2}(\displaystyle\frac{\pi km}{N})}\hspace{1.3cm} \\
   &  \\
   & <\displaystyle\frac{4\Gamma}{TN}\displaystyle\sum_{k=1}^{[N/2]}\displaystyle\frac{1-\exp[-\displaystyle\frac{T}{2\Gamma}\displaystyle\sum_{m=1}^{l}\sin^{2}(\displaystyle\frac{\pi
km}{N})]}{\displaystyle\sum_{m=1}^{l}\sin^{2}(\displaystyle\frac{\pi km}{N})}\hspace{1.7cm} \\
   \end{array}
   \end{eqnarray}
We assume that $T/\Gamma\gg1$ and obtain
\begin{eqnarray}\label{44}
\begin{array}{cc}
  \displaystyle\sum_{j=0}^{N-1}|\bar{P}(j,t)-\displaystyle\frac{1}{N}|  & <\displaystyle\frac{\Gamma}{T}\displaystyle\sum_{k=1}^{[N/2]}\displaystyle\frac{N^{2}}{k^{2}\displaystyle\sum_{m=1}^{l}m^{-2}}\hspace{1.5cm} \\
   &  \\
   & <\displaystyle\frac{\Gamma N^{2}}{T\displaystyle\sum_{m=1}^{l}m^{-2}}\zeta(2)\leq\displaystyle\frac{\Gamma N^{2}\pi^{2}}{6T\displaystyle\sum_{m=1}^{l}m^{-2}} \\
   \end{array}
   \end{eqnarray}
where $\zeta$ is Riemann zeta function~\cite{GAHW}.
\\Thus, the upper bound of average mixing time is
\begin{eqnarray}\label{35}
t_{ave,upper}=\displaystyle\frac{\Gamma
N^{2}\pi^{2}}{6\epsilon\displaystyle\sum_{m=1}^{l}m^{-2}},
   \end{eqnarray}
Therefore
\begin{eqnarray}\label{44}
\begin{array}{c}
  \displaystyle\frac{2\Gamma}{\sum_{m=1}^{l}
\sin^{2}(\pi m/N
)}\ln(\displaystyle\frac{2}{N\epsilon})<t_{ins}<\displaystyle\frac{\Gamma
N^{2}}{2\sum_{m=1}^{l}m^{-2}}\ln(\displaystyle\frac{2+\epsilon}{\epsilon})\\
   \\
  \displaystyle\frac{4\Gamma}{N\epsilon
\sum_{m=1}^{l}\sin^{2}(\pi m/N)}<t_{ave}<\displaystyle\frac{\Gamma
N^{2}\pi^{2}}{6\epsilon\sum_{m=1}^{l}m^{-2}}
\end{array}
   \end{eqnarray}

As we see, the mixing time bounds increase with increasing of the
rate of decoherence $\Gamma$. The reason being that when a
double-dot system having aligned levels of energy is strongly
measured by a point-contact detector, a quantum Zeno effect emerges.
According to Zeno effect, the strong measurement process slow down
transitions between quantum states due to the collapse of the wave
function into the observed state, which increases the localization
of electron in the initial dot and destroys the mixing
time\cite{DSLF, Gurvitz}. Also, comparing of the both mixing times
shows that the instantaneous mixing happens earlier than the
time-average mixing. Moreover, the mixing time bounds decrease with
adding the newly edges.
\section{Conclusion}
We considered the continuous-time quantum walks on one-dimension
regular network under large decoherence $\Gamma\gg 1$. For this, we
used an analytical model developed by Gurvitz~\cite{G} and
calculated the probability distribution. Then we obtained the lower
and upper bounds of instantaneous and average mixing times as
\begin{eqnarray}\label{47}\nonumber
  \displaystyle\frac{2\Gamma}{\sum_{m=1}^{l}
\sin^{2}(\pi m/N
)}\ln(\displaystyle\frac{2}{N\epsilon})<t_{ins}<\displaystyle\frac{\Gamma
N^{2}}{2\sum_{m=1}^{l}m^{-2}}\ln(\displaystyle\frac{2+\epsilon}{\epsilon})
\end{eqnarray}

\begin{eqnarray}\label{35}\nonumber
  \displaystyle\frac{4\Gamma}{N\epsilon
\sum_{m=1}^{l}\sin^{2}(\pi m/N)}<t_{ave}<\displaystyle\frac{\Gamma
N^{2}\pi^{2}}{6\epsilon\sum_{m=1}^{l}m^{-2}}
   \end{eqnarray}
Thus the instantaneous mixing time is shorter than the average one,
and the both mixing times are linearly proportional to the
decoherence rate. We found that adding shortcuts to cycle network
decreases the mixing times. Moreover, our analytical results for
$l=1$ are in agreement with the mentioned results in~\cite{FST}.


\begin{thebibliography}{99}
\bibitem{FLS} R.P. Feynman, R.B. Leighton and M. Sands, Feynman lectures on physics, Addison
Wesley, 1964.
\bibitem{EFSG} E. Farhi and S. Gutmann, Phys. Rev. A \textbf{58}, 915 (1998).
\bibitem{ADZ} A. Ambainis, E. Bach, A. Nayak, A. Vishwanath, and J. Watrous,
in Proceedings of the 33rd Annual ACM Symposium on Theory of
Computing (STOC'01) (ACM Press, New York, 2001), Pages: 37 - 49.
\bibitem{NK} N. Konno, Quantum Information Processing, Vol. \textbf{8}, No. 5, 387
(2009).
\bibitem{HKTB} H. Krovi and T.A. Brun, Phys. Rev. A \textbf{75}, 062332 (2007).
\bibitem{MB} O. M\"{u}lken and A. Blumen, Phys. Rev. A \textbf{73}, 012105 (2006).
\bibitem{CMC} C.M. Chandrashekar, arXiv: quant-ph/0609113V4 (2006).
\bibitem{Gottlieb} A. D. Gottlieb, Phys. Rev. E \textbf{72}, 047102
(2005).
\bibitem{tamon}D. Avraham, E. Bollt and C. Tamon, Quantum Information Processing \textbf{3}, 295
(2004).
\bibitem{SSS} S. Salimi, Annals of Physics \textbf{324}, Pages: 1185-1193
(2009).
\bibitem{xp1} X. Xu, J. Phys. A: Math. Theor. \textbf{42}, 115205 (2009).
\bibitem{SJDP} S. Salimi and M. Jafarizadeh, Commun. Theor. Phys.
\textbf{51}, Pages: 1003-1009 (2009).
\bibitem{SSQG} S. Salimi, Quantum Information Processing, Vol. \textbf{6}, 945 (2008).
\bibitem{SSOG} S. Salimi, Int. J. Theor. Phys. \textbf{47}, Pages: 3298-3309 (2008).
\bibitem{saja} M.A. Jafarizadeh, S. Salimi, Ann. Phys. \textbf{322}, 1005
(2007).
\bibitem{NKT} N. Konno, Infinite Dimensional Analysis, Quantum
Probability and Related Topics, Vol. \textbf{9}, No. 2, Pages:
287-297 (2006).
\bibitem{XPXR1} X. Xu, Phys. Rev. E \textbf{79}, 011117 (2009).
\bibitem{NKUS} N. Konno, International Journal of Quantum
Information, Vol. \textbf{4}, No. 6, Pages: 1023-1035 (2006).
\bibitem{DHS} M. Drezgi\'{c}, A. P. Hines, M. Sarovar and Sh.
Sastry, Quantum Information and Comp. \textbf{9}, 854 (2009).
\bibitem{Kendon} V. Kendon, Math. Struct. in Comp. Sci \textbf{17}, No. 6,
1169(2006).
\bibitem{FWS} F. W. Strauch, quant-ph/0808.3403 (2008).
\bibitem{AR} G. Alagic and A. Russell, Phys. Rev A \textbf{72},
062304 (2005).
\bibitem{RSAAD} A. Romanelli, R. Siri, G. Abal, A. Auyuanet and R.
Donangelo, J. Phys. A, Vol. \textbf{347C}. Pages: 137-152 (2005).
\bibitem{KTNC} V. Kendon, B. Tregenna, Phy. Rev. A \textbf{67}, 042315 (2003).
\bibitem{SRLS} S. Salimi and R. Radgohar, J. Phys. A: Math. Theor. \textbf{42}, 475302 (2009).
\bibitem{SRLL} S. Salimi and R. Radgohar, J. Phys. B: At. Mol. Opt. Phys. \textbf{43}, 025503
(2010).
\bibitem{FST} L. Fedichkin, D. Solenov and C. Tamon, Quantum Information and
Computation, Vol 6, No. \textbf{3} ,Pages: 263-276 (2006).
\bibitem{Dur} W. D\"{u}r, Phys. Rev. A \textbf{66}, 052319 (2002).
\bibitem{cote} R. C\^{o}t\'{e}, New J. Phys. \textbf{8} 156 (2006).
\bibitem{SRR} S. Salimi and R. Radgohar, International Journal of Quantum Information Vol. \textbf{8}, No. 5,  795 (2010).
\bibitem{XPXR} X. Xu, Phys. Rev. E \textbf{77}, 061127 (2008).
\bibitem{HSIS} S.H. Strogatz, and I. Stewart, Sci. Am. \textbf{269}, 102 (1993).
\bibitem{KW} K. Wiesenfeld, Physica B \textbf{222}, 315 (1996).
\bibitem{IBBH} I.V. Belykh, V.N. Belykh and M. Hasler, Physica D \textbf{195}, 159 (2004).
\bibitem{DWHS} D.J. Watts and S. H. Strogatz, Nature \textbf{393}, 440 (1998).
\bibitem{CG} A.M. Childs and J. Goldstone, Phys. Rev. A \textbf{70},
022314-022324 (2004).
\bibitem{MB5} O. M\"{u}lken and A. Blumen, Phys. Rev. E \textbf{71},
016101-016106 (2005).
\bibitem{V} A. Volta, O. M\"{u}lken and A. Blumen, J. Phys. A \textbf{39},
14997-15012 (2006).
\bibitem{MW} E.W. Montroll and G. H. Weiss, J. Math. Phys. \textbf{6},
167-181 (1965).
\bibitem{CFG} A.M. Childs, E. Farhi and S. Gutmann, Quantum
Information Processing \textbf{1}, 35 (2002).
\bibitem{OMAB} O. M\"{u}lken and A. Blumen, Phys. Rev. E \textbf{71}, 036128 (2005).
\bibitem{JMZ} J.M. Ziman, Principles of the Theory of Solids (Cambridge
University Press, Cambridge, England, 1972).
\bibitem{DSLF} Dmitry Solenov and Leonid Fedichkin, "Continuous-Time Quantum Walks on a Cycle
Graph," Phys. Rev. A \textbf{73}, 012313 (2006).
\bibitem{torre} A.C. de la Torre, H. O. Mártin and D.
Goyeneche, Phys. Rev. E \textbf{68}, 031103 (2003).
\bibitem{Pioro} M. Pioro-Ladriere, R. Abolfath, P. Zawadzki, J.
Lapointe, S. A. Studenikin, A. S. Sachrajda  and P. Hawrylak, Phys.
Rev. B \textbf{72}, 125307 (2005).
\bibitem{APS} A.P. Hines and P.C.E. Stamp, quant-ph/0701088
(2007).
\bibitem{G} S.A. Gurvitz, Phys. Rev. B \textbf{57}, 6602 (1998); S.
A. Gurvitz, Phys. Rev. B \textbf{56}, 15215 (1997).
\bibitem{MASH} M. Abramowitz and I.A. Stegun, Handbook of Mathematical Functions, Dover (1972);
\bibitem{AJVK} D. Aharonov, A. Ambainis, J. Kempe and U. Vazirani, Proceedings of ACM Symposium on Theory of Computation (STOC 01), Pages: 50-59 (2001).
\bibitem{GAHW} G.B. Arfken and H.J. Weber, Mathematical Methods for
Physicists, Chapter 5, Harcourt Academic Press (2005).
\bibitem{Gurvitz} S.A. Gurvitz, Quantum Information
Processing, Vol. \textbf{2}, 15 (2003).




\end{thebibliography}
\end{document}